\documentclass[useAMS,usenatbib]{mn2e}
\usepackage{epsfig}

\def\msun{{\rm {M}}_{\odot}}
\newcommand{\etal}{{et al.}~}
\newcommand{\eg}{{e.g.~}}
\def \ltsima{$\; \buildrel < \over \sim \;$}
\def \simlt{\lower.5ex\hbox{\ltsima}}            % < over ~
\def \gtsima{$\; \buildrel > \over \sim \;$}
\def \gtsima{\mbox{$\; \buildrel > \over \sim \;$}}
\def \simgt{\lower.5ex\hbox{\gtsima}}            % > over ~
\title[Evolution of the Mass Function]{Evolution of the Mass Function of Dark Matter Haloes}
\author[Reed \etal] {Darren Reed,$^1$\thanks{Email: reed@astro.washington.edu} Jeffrey Gardner,$^2$ 
Thomas Quinn,$^1$
\newauthor
Joachim Stadel,$^3$ Mark Fardal,$^4$ George Lake,$^1$
\newauthor
and Fabio Governato$^{1,5}$\\
$^1$Astronomy Department, University of Washington, Seattle, USA\\
$^2$Phys. and Astronomy Department, University of Pittsburg, USA\\
$^3$Institute for Theoretical Physics, University of Zurich, Switzerland\\
$^4$Astronomy Department, University of Victoria, Canada\\
$^5$Osservatorio Astronomico di Brera, Milano, Italy}

\pagerange{\pageref{firstpage}--\pageref{lastpage}}
\pubyear{2003}

\begin{document}

\maketitle

\label{firstpage}

\begin{abstract}
We use a high resolution $\Lambda$CDM numerical simulation to calculate the mass function
of dark matter haloes down to the scale of dwarf galaxies, back to a redshift of
fifteen, in a 50 $h^{-1}$Mpc volume containing 80 million particles.  
Our low redshift results allow us to probe low $\sigma$ density fluctuations
significantly beyond the range of previous cosmological simulations.
The Sheth and Tormen mass function provides an excellent match to all of our data except 
for redshifts of ten and higher, where it
overpredicts halo numbers increasingly with redshift, reaching roughly 50 percent for the 
$10^{10}-10^{11} \msun$ haloes sampled at redshift 15.  
Our results confirm previous findings that the simulated halo
mass function can be described solely by the variance of the mass distribution, 
and thus has no explicit redshift dependence.
We provide an empirical fit to our data that corrects for the
overprediction of extremely rare objects by the Sheth and Tormen mass function.
This overprediction has implications for studies that use the number densities
of similarly rare objects as cosmological probes.  For example, the number density
of high redshift (z $\simeq$ 6) QSOs, which are thought to be hosted
by haloes at 5$\sigma$ peaks in the fluctuation field, are likely to be 
overpredicted by at least a factor of 50$\%$.  We test the sensitivity of our 
results to force accuracy, starting redshift, and halo finding algorithm.

\end{abstract}

\begin{keywords} galaxies: haloes -- galaxies: formation -- galaxies:
clustering -- cosmology: theory -- cosmology:dark matter
\end{keywords}

\section{Introduction}

Cold dark matter models with a cosmological constant ($\Lambda$CDM)
are successful in explaining a wide array of kinematic and structural
properties of the observed universe.  A critical test of the
$\Lambda$CDM model is how well it predicts the abundance of dark
matter haloes, which serve as hosts for observable clusters, groups,
and galaxies.  Simulations that resolve haloes out to high redshift
can be used to model the evolution of the numbers of observable high
redshift objects, their progenitors, and their evolved descendants.
Lyman break galaxies, for example, observed at redshifts out to z
$\simeq$ 4 (\eg Steidel \etal 1996) are likely progenitors of groups
or clusters (Governato  \etal 1998; Governato \etal 2001), and we are
able to model their numbers over their entire observable lifespan.
Many objects, however, lie outside the realm that can presently be
simulated because their number densities, masses, or redshifts are too
extreme.  For example, to model a reasonably sized sample of the
hosts of the highest redshift (z $\simeq$ 6) quasi-stellar objects
(QSOs), whose number density has been measured by the Sloan Digital
Sky Survey (Fan \etal 2001), would require a simulation with volume
roughly as large as the observable universe with (so far)
prohibitively high particle numbers.   Simulations of adequate
resolution and volume could be used, in principle, to estimate the
host masses of such rare QSOs, or  to estimate cosmological parameters
after assuming host masses, by matching  predicted and observed number
densities.  However, with much less computational effort, by modeling
smaller cosmological volumes at higher redshift, we are able to test
analytic mass functions in the same regime of rare density
enhancements.  Generally, rare density peaks correspond to high values
of M/M$_*$, where M$_*$ is the `characteristic' mass of a typical
collapsing halo at that epoch (to be discussed later).  By modeling
haloes over a wide range of M/M$_*$, we can constrain analytic mass
functions for ranges of redshift and mass that have not yet been
simulated.  This is possible because analytic mass functions are
generally derived from assumptions of how linear density fluctuations
lead to halo collapse, and thereby have no explicit redshift
dependence.  High redshift QSO hosts, galaxy progenitors, and perhaps
even the first generation of stars are all examples of high $M/M_*$
objects whose mass functions can presently only be calculated
analytically.

Press \& Schechter (P-S, 1974) developed an analytical framework that
predicts the number and formation epoch of dark matter haloes.  In P-S
theory, as the universe evolves, linear density fluctuations grow
gravitationally until they reach a critical spherical overdensity, at
which time non-linear gravitational collapse is assumed to occur.
Cosmological numerical simulations have shown the P-S framework to be
approximately correct, but P-S theory consistently underpredicts the
number of high mass haloes and overestimates the number of haloes less
than about $M_*$ (\eg Efstathiou \etal 1988; Gross \etal 1998;
Governato \etal 1999; Jenkins \etal 2001; White 2002) even when
merging of dark matter haloes is included in predictions (Bond \etal
1991; Bower 1991; Lacey \& Cole 1993; Gardner 2001), though the high
mass end fits  well if the finite size of haloes is taken into account
(Yano, Nagashima, \& Gouda 1996; Nagashima 2001).  Ellipsoidal halo
collapse models (\eg Monaco 1997ab; Lee \& Shandarin 1998; Sheth, Mo,
\& Tormen 2001) yield much more robust predictions than the
conventional spherical collapse models, and are in excellent agreement
with empirical fits by Sheth \& Tormen (1999).  Monaco \etal (2001),
using the semi-analytic code PINOCCHIO, which uses a perturbative
approach, show that the dark matter halo distribution can be
accurately predicted at much lower computational cost, on a
point-by-point basis, from a numerical realization of an initial
density field.  Jenkins \etal (2001) utilize a large set of
simulations of a range of volumes and cosmologies (Jenkins \etal
1998; Governato \etal 1999; Evrard \etal 2002) to test the Sheth \&
Tormen (S-T) mass function over more than four orders of magnitude in
mass, and out to a redshift of 5, finding good agreement with the S-T
function down to their resolution limit of $\simeq 3 \times 10^{11}
\msun$, except for an overprediction by the S-T function for haloes
at rare density enhancements.   In this study, we
probe previously untested regimes of the mass function by simulating a
volume that resolves haloes down to the scale of $10^{10}\msun$ 
dwarfs, in a
cosmological environment, allowing us to sample the mass function back
to z$\simeq$15.
Our paper is outlined as follows: in section 2, we describe our
simulations and numerical techniques; in section 3, we review the
analytic theory; then we discuss our results and compare with previous
work in section 4; we conclude with a discussion of the implications
of our work.

\section{The simulations}

Our cosmology is the presently favored $\Lambda$CDM with $\Lambda =
0.7$ and $\Omega_{m} = 0.3$.  We use the parallel tree gravity solver
PKDGRAV (Stadel 2001) to simulate $81 \times 10^{6}$ (432$^3$) dark
matter particles from a starting redshift, z$_0$, of 69.  We then
resimulate the same volume but with z$_0=$139, and evolve this volume
to z$=$7; which allows us to consider results at higher redshift than
with the  z$_0=$69 run.  In order to simulate the highest possible
mass resolution, we employ a volume of 50 $h^{-1}$Mpc on a side.  Our
particle mass is $1.3 \times 10^{8} h^{-1}\msun$ allowing us to
resolve haloes down to less than $10^{10} h^{-1}\msun$ with 75
particles.   Our force resolution is 5 $h^{-1}$kpc.  We use a cell
opening angle of $\Theta <$ 0.8 at low redshift, and $\Theta <$ 0.7 at
z $>$ 2.  In section 4.2, we discuss tests that confirm that our
choices of initial redshift,  softening, and high redshift opening
angle are adequate (see Table 1).  We use a ``multistepping''
approach, where particles in the highest density regions undergo
16,000 timesteps.  Timesteps were constrained to $\delta t < 0.2
\sqrt{\epsilon/a}$, where $\epsilon$ is the softening length and $a$
is the magnitude of the acceleration of a given particle.  We
normalize the density power spectrum of our initial conditions  such
that $\sigma_{8}$, the rms density fluctuation of spheres of 8
$h^{-1}$Mpc extrapolated to redshift of zero is 1.0, consistent with
both the cluster abundance (see \eg Eke, Cole, \& Frenk 1996 and
references therein) and the COBE normalization (\eg Ratra \etal 1997).
To set our initial conditions, we use the Bardeen \etal (1986)
transfer function with $\gamma=\Omega_{m0} h$, where $h$ is the hubble
constant in units of 100 $km s^{-1} Mpc{-1}$.

\subsection{Halo Identification}

In order to identify haloes in our simulation, we use both the {\it
friends-of-friends} (FOF) algorithm (Davis \etal 1985), and the {\it
spherical overdensity} (SO) algorithm (Lacey \& Cole 1994).  The FOF
halo finder uses a ``linking length'', {\it ll}, to link together all
neighboring particles with spacing closer than {\it ll} as members of
a halo.  SO identifies haloes by identifying spherical regions with
the expected spherical overdensities of virialized haloes.  For our SO
haloes, we first use SKID (Stadel 2001) to identify all bound haloes,
including those that are subhaloes within larger haloes.  Next, we
grow a sphere outward from each SKID center until it just contains the
virialized overdensity.  Finally, we iterate by growing spheres
outward from all neighboring SKID haloes until we have identified the
center of mass for each SO halo.  For our SO criterion, we use the
virial overdensity predicted by the spherical collapse tophat model of
Eke, Cole, \& Frenk (1996).  In a $\Lambda$CDM universe, the virial
overdensity, in units of critical density, declines from an asymptotic
value of 178 as cosmic time increases and $\Omega_{m}$ drops below 1;
for redshift of zero, the $\Lambda$CDM overdensity {\it
$\Delta_{vir}$} is 100 (Kitayama \& Suto 1996).  We exclude haloes
that contain less than 64 particles, which is conservative, as it is
more than the estimated 20 or 30 particles needed for a robust halo
identification based on  resolution tests (Jenkins \etal 2001;
Governato \etal 1999).

We utilize the FOF algorithm for the bulk of our analyses since it is
reasonably robust and computationally efficient.  Our FOF {\it ll}
choice is 0.2 for all redshifts (except when matching $ll$ from
previous studies); this approach has been shown to be sound for a
range of cosmologies by Jenkins \etal (2001), although the evolution
of {\it $\Delta_{vir}$} in the spherical collapse tophat model implies
that $ll$ should range from $ll=$0.164 at z$=$0 to $ll=$0.2 at high
redshift in $\Lambda$CDM cosmology (Lacey \& Cole 1994; Eke, Cole, \&
Frenk 1996; Jenkins \etal 2001).  To test the sensitivity of the mass
function to halo selection criteria, we apply both FOF $ll=$0.164 and
SO to our data at various redshifts, discussed in section 4.1.

\section{Analytic Theory}

Analytic P-S formalism yields the following (See Jenkins \etal 2001,
whose notation we  adopt in this section, and references therein for
further discussion, which we summarize here):
\begin{equation}\label{ps_massfn}
   f(\sigma; {\rm P{\rm-}S}) = \sqrt{2\over\pi}
   {\delta_c\over\sigma}\exp\bigg[-{\delta_c^2\over2\sigma^2}\bigg],
\end{equation}
where $\delta_c$ is the threshold spherical linear overdensity above
which a region will collapse.  $\delta_c$ depends only weakly on the
cosmological parameters and redshift (\eg More, Heavens, \& Peacock
1986; Jenkins \etal 2001), and $\delta_c = 1.686$ for $\Omega_0 = 1$.
In our analysis, we assume that $\delta_c = 1.686$ at all redshifts.
$\sigma^2(M,z)$ is the variance of the linear density field, smoothed
with a spherical top-hat filter enclosing mass $M$, and is calculated
from the linear density power spectrum $P(k)$, extrapolated to $z=0$:
\begin{equation}\label{vardeff}
\sigma^2(M,z)  =  {b^{2}(z)\over2\pi^2}\int_0^\infty
k^2P(k)W^2(k;M){\rm d}k,
\end{equation}
where $W(k;M)$ is the Fourier-space top-hat filter, and ${\it b(z)}$
is the growth factor of linear perturbations normalized to unity at
z$=$0 (Peeble 1993).  In the P-S formalism, all mass is contained in
haloes:
\begin{equation}\label{integcond}
   \int_{-\infty}^\infty f(\sigma; {\rm P{\rm-}S}){\rm
   d}\ln\sigma^{-1} = 1.
\end{equation}
The mass function $f(\sigma, z)$ can be related to the number density,
$n(M, z)$, of haloes with mass less than $M$:
\begin{equation}\label{deff}
   f(\sigma, z) \equiv \frac{M}{\rho_0(z)}{{\rm d}n(M, z)\over{\rm
d}\ln\sigma^{-1}},
\end{equation}
where $\rho_0(z)$ is the mean density of the universe at that time.

The S-T model is a modification to the P-S model based on empirical
fits to simulations (Sheth \& Tormen 1999), has been shown to
reproduce simulation results substantially better than P-S (\eg
Jenkins \etal 2001; White 2002), and is theoretically justified in
that it matches P-S formalism derived with ellipsoidal halo collapse
models (Sheth, Mo, \& Tormen 2001):
\begin{equation}\label{sheth_tormen}
  f(\sigma; {\rm S{\rm-}T}) = A\sqrt{{2a\over\pi}}
\bigg[1+\big({\sigma^2\over a\delta_c^2}\big)^p\bigg]
{\delta_c\over\sigma}\exp\bigg[-{a\delta_c^2\over2\sigma^2}\bigg],
\end{equation}
where A=0.3222, $a=0.707$ and $p=0.3$.  Jenkins \etal (2001) offer an
empirical fit using high resolution simulations of a range of
cosmologies.  Their fit is constructed in the $f - \ln(\sigma^{-1})$
plane, which has the advantage of being invariant with redshift:
\begin{equation}
\label{tcdm_fit2}
 \phantom{xxxxxxx}f(\ln\sigma^{-1}) =
 0.315\;\exp\big[-|\ln\sigma^{-1}+0.61|^{3.8}\big].
\end{equation}
The Jenkins \etal function adjusts for an overprediction by the S-T
function for the rare objects at large $\ln\sigma^{-1}$, and is
calibrated for the range $-1.2\le\ln\sigma^{-1}\le1.05$, which
corresponds to masses down to approximately 3$\times10^{11}
h^{-1}\msun$ at present epoch, and includes haloes out to z$=$5, with
FOF fixed at $ll=$0.2.

In each of these analytic functions, virialized haloes have a
characteristic mass, {\it $M_*$(z)}:
\begin{equation} \sigma(M_*(z)) = \delta_c, 
\end{equation}
and $\sigma(M,z) = \sigma(M,z=0)b(z)$.  ${\it b(z)}$ evolves as
$(1+z)^{-1}$ in an $\Omega_0 = 1$ universe, and more slowly in a
$\Lambda$CDM universe.  $\sigma(M)$ decreases slowly with increasing
mass, which leads to the steep redshift dependence of $M_*$(z), shown
in Fig. 1 for the $\Lambda$CDM universe, and results in a broad mass
spectrum of collapsed objects.  
An important test of analytic mass functions is the accuracy of
their predictions at high values of $M/M_*$ and $z$.
At low redshift, high $M/M_*$ haloes would have
unrealistically large masses, but at high redshift the steep evolution
of $M_*$ puts high $M/M_*$ objects well into the realm of simulations.
Furthermore, the evolution of halo masses that correspond to high $\sigma$ 
density enhancements
means that rare haloes are most easily simulated at 
high redshift (Fig. 1).  Our simulations model the mass function of
haloes lying at up to 4$\sigma$ density fluctuations.

\begin{figure}
\begin{center}
\epsfig{file=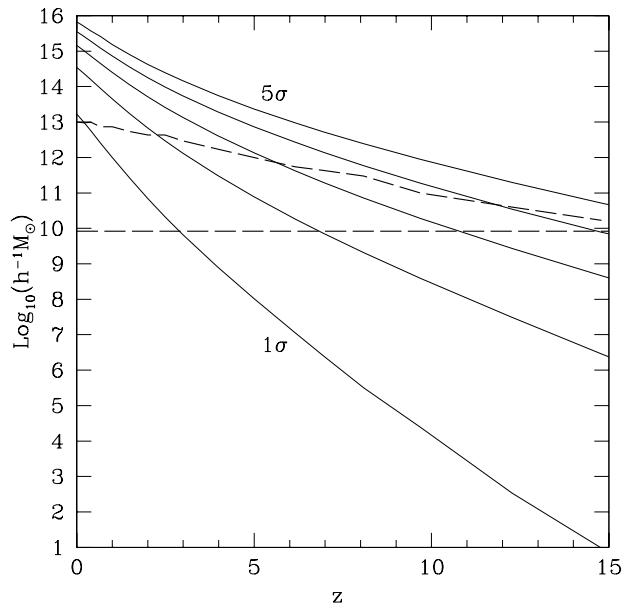, width=\hsize}
\caption{The curves correspond to the halo mass of $n\sigma$ fluctuations in the density
field, given by $\sigma(M, z) = \delta_c/n$, with $n=$1, 2, 3, 4, and 5, from bottom to top.  
The characteristic mass, $M_{*}$, given by $\sigma(M_*(z)) = \delta_c$, is the 1$\sigma$ curve.
The area between the long dashed lines is sampled with poisson errors of less than
20$\%$ in our data.}
\label{fig:mstar}
\end{center}
\end{figure}

\section{Evolution of the Mass Function}
In Fig. 2, we compare the analytic version of the P-S, the S-T, and
the Jenkins \etal mass function with our simulation results at several
redshifts.  The S-T function provides the best fit to our simulation
with excellent agreement at all masses and redshifts except for our
highest redshift outputs.  The P-S function overpredicts substantially
everywhere except at the low and high mass extremes.  And the Jenkins
\etal mass function fits much of our data well at z$=$0, but
diverges from our simulation results once well below the
%%%3$\times10^{11} h^{-1}\msun$ 
limit of its empirical fit of $\ln\sigma^{-1}=-1.2$, 
which corresponds to $\simeq4\times10^{11} h^{-1}\msun$ with $\sigma_{8}=1.0$.
Note that $\sigma(M,z)$ is sensitive to $\sigma_{8}$, so 
for a $\sigma_{8}=0.9$ $\Lambda$CDM model, which was used in many of the
simulations that were part of the Jenkins \etal fit, $\ln\sigma^{-1}=-1.2$
would correspond to only $\simeq2\times10^{11} h^{-1}\msun$.

\begin{figure}
\begin{center}
\label{fig:nsubm}
\epsfig{file=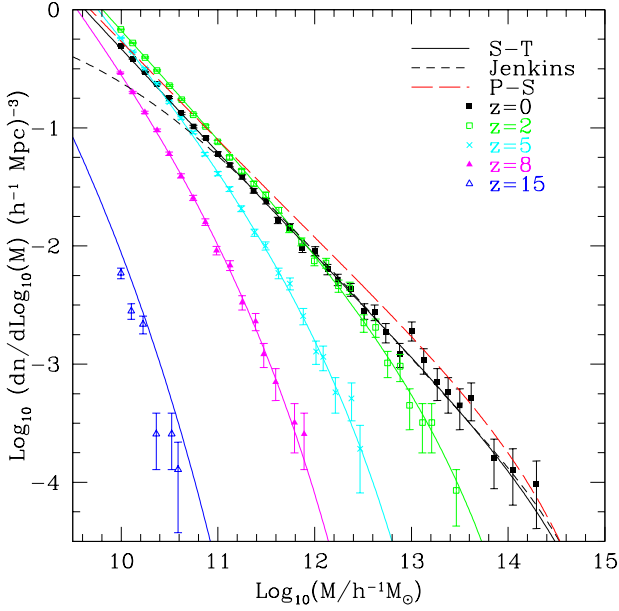, width=\hsize}
\caption{Comparison of the mass function per decade of mass.  Data
points are our $\Lambda$CDM simulation results with 1$\sigma$ poisson
error bars.  Throughout the  paper, haloes are identified using a FOF
$ll=$0.2, unless otherwise specified; only haloes with at least 64
particles are considered.  In our plots, we plot the median halo mass
in a bin.  Solid curves are the Sheth \& Tormen function at z$=$0, 2,
5, 8, \& 15.  Short dashed curve is the Jenkins \etal ``universal''
mass function (Eqn. 6), which diverges when extrapolated
well below its original lower mass limit of $\simeq4\times10^{11} h^{-1}\msun$
for our $\sigma_8=1.0$ model (see text).  
Long dashed curve is the Press \& Schechter function.}
\end{center}
\end{figure}

In Fig. 3, we show the evolution of the mass function over all of our
redshifts compared to its S-T predicted evolution.  The S-T function
provides an excellent fit to our data, except at very high redshifts,
where it significantly overpredicts the halo abundance.  At all
redshifts up to z$=$10, the difference is $\simlt 10 \%$ for each of
our well sampled mass bins.  However, the S-T  function begins to
overpredict the number of haloes increasingly with redshift for
z$\simgt$10, up to $\sim 50$\% by z$=$15.  The simulation mass
functions appear to be generally steeper than the S-T function,
especially at high redshifts.  In Fig. 4, we show the evolution of the
mass function over all of our redshifts as a function of $M/M_*$.
This highlights the remarkable accuracy of the S-T mass function over
more than 10 decades of $M/M_*$.

\begin{figure}
\label{fig:tilediff}
\begin{center}
\epsfig{file=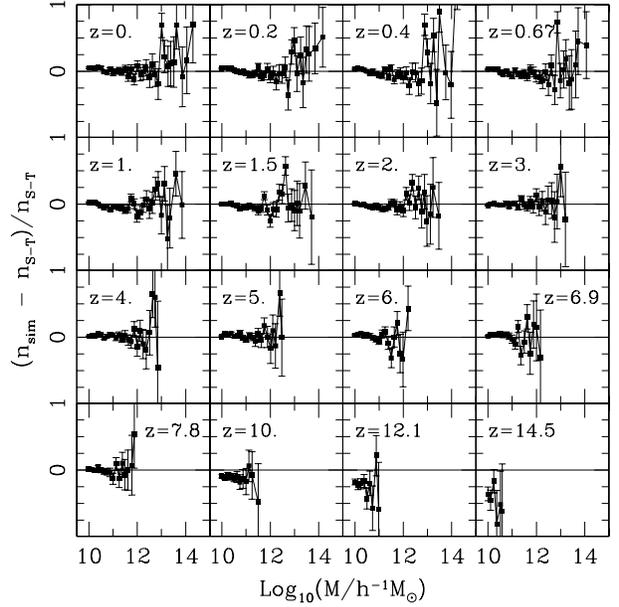, angle=0, width=\hsize}
\caption{Fractional difference between our simulated mass function
(FOF $ll=$0.2) and the S-T prediction.  Poisson error bars shown.  }
\end{center}
\end{figure}

\begin{figure}
\begin{center}
\label{fig:tilemstar}
\epsfig{file=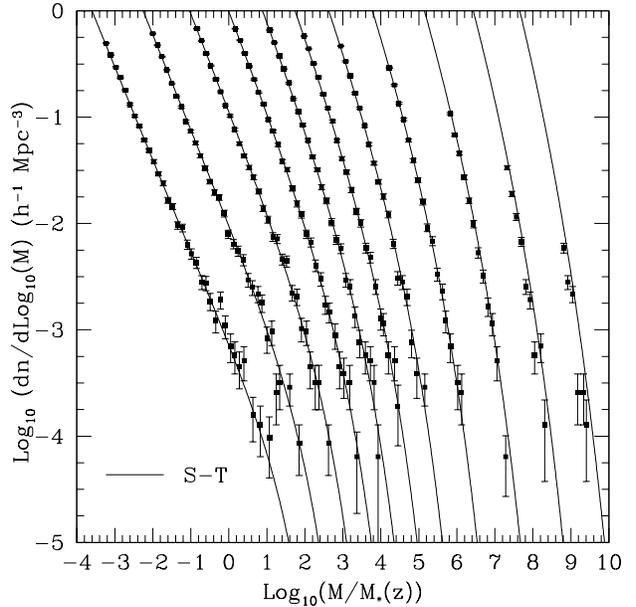, angle=0, width=\hsize}
\caption{Number density vs. M/M$^*$.  Data points with poisson error
bars are our simulation results.  Curves are S-T predictions.
Redshifts plotted are (from left to right) 0, 1., 2., 3., 4., 5., 6.2,
7.8, 10., 12.1, 14.5}
\end{center}
\end{figure}

In Fig. 5, we plot the mass function for all of our outputs in the $f
- \ln(\sigma^{-1})$ plane.  Large values of $\ln\sigma^{-1}$
correspond to rare haloes of high redshift and/or high  mass, while
small values of $\ln\sigma^{-1}$ describe haloes of low mass and
redshift  combinations.  In Fig. 6, we compare our simulated mass
function with the S-T prediction, by plotting the residuals over our
entire range.  We limit plotted data to bins with poisson errors of
less than 20\%.  Remarkably, the S-T function fits our simulated mass
function to better than 10$\%$ over the range of -1.7 $\leq
\ln\sigma^{-1} \leq$ 0.5.  We are unaware of any previous studies that
probe the mass function down to such low values of $\ln\sigma^{-1}$ in
a cosmological environment.  The S-T function appears to significantly
overpredict haloes for $\ln\sigma^{-1} \geq$ 0.5.  This is the same
overprediction seen in the number density for $z\geq 10$ in figs.  3
and 4.  The large apparent scatter of the mass function for
ln$\sigma^{-1}\simgt$ 0.5 is due to larger poisson errors in this
range.  For large ln$\sigma^{-1}$, we estimate the uncertainty in the
mass function due to cosmic variance by estimating the contribution of
linear fluctuations  on the scale of the box size.  Cosmic variance
can have large effects on results since it has the potential to
increase or decrease the mass function  in multiple mass bins
simultaneously, and cosmic variance is difficult to quantify without a
large set of simulated volumes.  In our estimation, we use a second
order  Taylor expansion of $f(\sigma; {\rm S{\rm-}T})$, ignoring the
first order term since we have imposed the average density of our
simulation to be $\Omega_m$.  The resulting estimate for the
uncertainty in the mass function due to cosmic variance is $\Delta_{f,
c.v.} \sim {{\partial^{2} f(\sigma; {\rm
S{\rm-}T})}\over\partial\sigma^{2}}{1\over{2}}{\sigma_{Mbox}^{2}}$,
which we  evaluate numerically.  In our well sampled, low redshift
mass bins, $\Delta_{f, c.v.}$ is negligible.  For our highest redshift
results, $\Delta_{f, c.v.}$ is smaller than our  poisson error limit
of 20$\%$ for figures 6 and 7, even for our highest mass bins.
However, $\Delta_{f, c.v.}$ approaches our poisson error limit of
20$\%$ for our $z=$14.5 output.  For our $z=$10 and $z=$12 outputs,
$\Delta_{f, c.v.}$ is less than 10$\%$ in bins where  poisson errors
are less than 10$\%$, which is the case for most of our bins at that
redshift.  Thus, while cosmic variance is a significant source of
error where the mass function is steepest, it is unlikely to entirely
account for our discrepancy with the S-T function.  We note that
several of our $z<2$ points lie roughly 3$\sigma$ above the mean; this
is actually just one mass bin plotted repeatedly at different
redshifts, and so is not entirely surprising.   By careful examination
of ranges of ln$\sigma^{-1}$ where outputs of different redshifts
overlap, we verify that the magnitude of the S-T overprediction at
high values of ln$\sigma^{-1}$ is consistent with being a function
purely of ln$\sigma^{-1}$ rather than redshift, a natural consequence
of the fact that  the mass function is self similar in time (\eg
Efstathiou \etal 1988; Lacey \& Cole 1994; Jenkins \etal 2001).
Jenkins \etal (2001) also find an overprediction by the S-T function
for $\ln\sigma^{-1} \simgt 0.75$, which with their larger simulation
volumes, corresponded primarily to objects of
z$\leq$2 and of much higher mass.  Additionally, Jenkins \etal find the mass function
to be invariant with redshift within their own results.  In Fig. 7, we
compare a subset of our data with the Jenkins \etal ``universal'' mass
function.   We note that when extrapolated to $\ln\sigma^{-1} \leq -1.4$, 
well below its empirical fit range of $-1.2\le\ln\sigma^{-1}\le1.05$,
the Jenkins \etal function diverges from our
results, reflecting the fact that it is of a
form not ideally suited for extrapolation.  Where our data overlap
($\ln\sigma^{-1} \geq -1.2$), we find generally good agreement,
although we have $\sim20\%$ fewer haloes and a somewhat steeper mass
function at $\ln\sigma^{-1} \simgt 0.25$.  Over the range of 
-1.4 $\simlt \ln\sigma^{-1} \simlt$ 0.75, the Jenkins \etal mass function
matches our data to within $\simeq20\%$.

We consider the possibility that this difference between our data  and
the Jenkins \etal fit could be due to  differences in the effective
slope of the power spectrum, n$_{eff}$, where $P(k) \propto
k^{n_{eff}}$.  Applying a simple power law to Eqn. 2 yields $\sigma^2
\propto M^{-(n_{eff}+3)/3}$, which can be reparameterized as
\begin{equation}
      n_{\rm eff} = 6 {{\rm d}\ln\sigma^{-1}\over{\rm d}\ln
      M\phantom{+}} -3,
\label{defnstar}\end{equation}
(Jenkins \etal 2001).  Fig. 8 shows the $n_{eff}~
vs. ~\ln\sigma^{-1}$ parameter space that our results cover.  Our data
generally has a steeper $n_{eff}$ than Jenkins et al., though there is
some overlap, especially at lower redshifts.  Since the slope of the
linear power spectrum is invariant with redshift for a given $k$,
n$_{eff}(z)$ is constant for a given mass, meaning that for our lowest
mass haloes we sample nearly all of our $f(\sigma)$ at an
n$_{eff}\simeq-2.3$.  If we consider $n_{eff}(\sigma=0.5)$, where the
power spectrum begins to go  nonlinear, then we find that our results
differ significantly from the Jenkins  \etal function only where
$n_{eff}$ at the nonlinear scale is steepest.  In  particular, the
disagreement is worst at z$\geq$10 where $n_{eff}(\sigma=0.5)$ exceeds
their maximum value of -2.26.  Thus, there is a possibility that our
steeper values of n$_{eff}$ results in an $f(\sigma)$ with a slightly
different shape, which might account for the difference with prior
results, but a larger set of simulations with still steeper $n_{eff}$
would be needed to clearly show any such dependence.

\begin{figure}
\begin{center}
\label{fig:stbig}
\epsfig{file=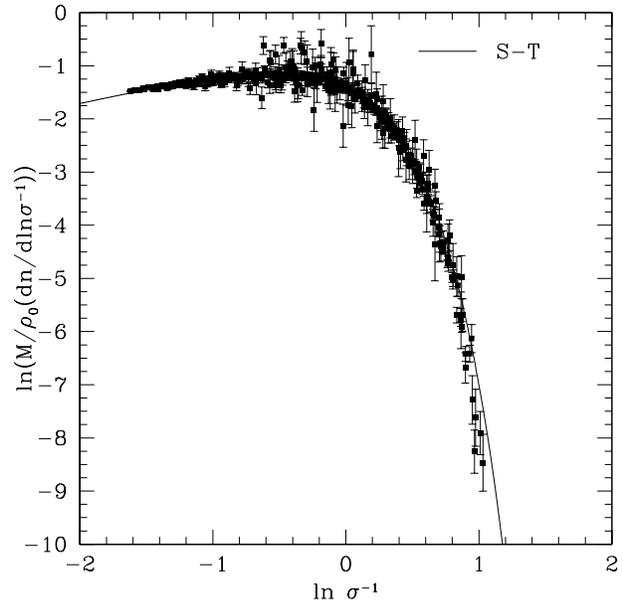, angle=0, width=\hsize}
\caption{Mass function plotted in redshift independent form for all of
our outputs as in Fig. 4.  Solid curve is S-T prediction.}
\end{center}
\end{figure}

\begin{figure}
\begin{center}
\label{fig:stdiff}
\epsfig{file=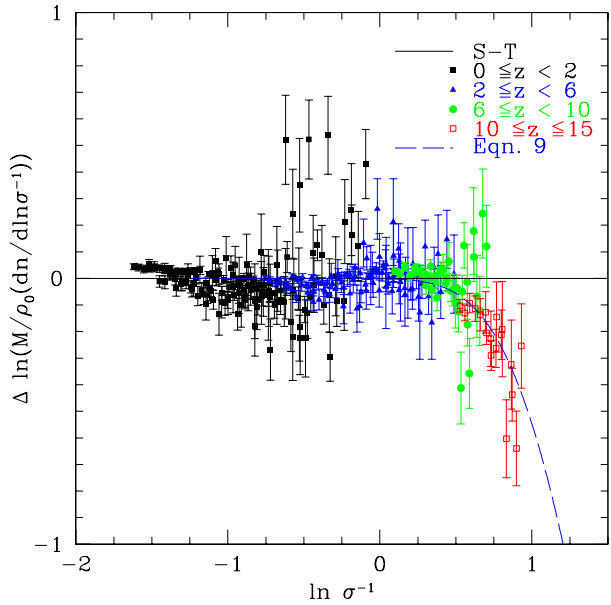, angle=0, width=\hsize}
\caption{Residuals between S-T prediction and our results for the mass
function of Fig. 5 (with FOF $ll$=0.2).  Solid straight line is the
S-T function.  Dashed line is our empirical adjustment to the S-T
function.  }
\end{center}
\end{figure}

\begin{figure}
\begin{center}
\label{fig:jenkdiff}
\epsfig{file=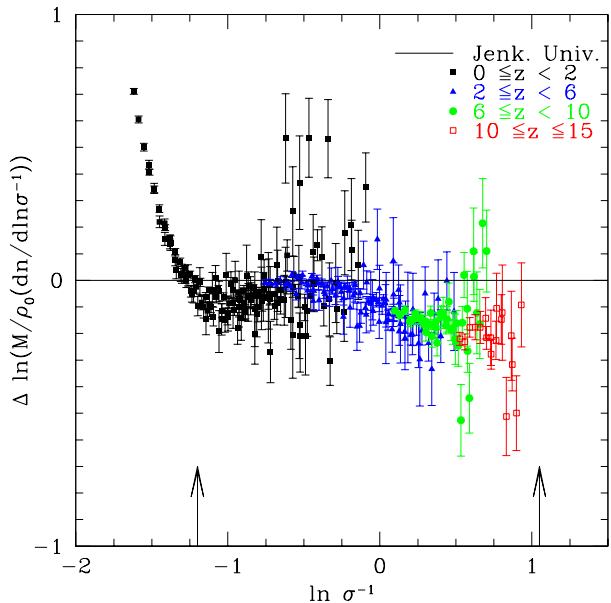, angle=0, width=\hsize}
\caption{ Residuals between Jenkins \etal mass function and our
results for the mass function of Fig. 5.  Arrows encompass the range
of data used in  the Jenkins \etal empirical fit, which is denoted by
the solid straight line.  }
\end{center}
\end{figure}

\begin{figure}
\begin{center}
\label{fig:neff}
\epsfig{file=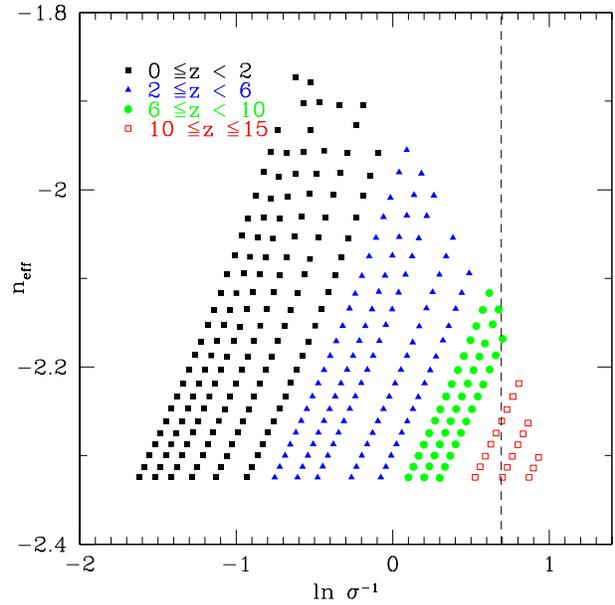, angle=0, width=\hsize}
\caption{Parameter range covered by our results.  Each mass bin for
which we have poisson statistics of better than 20$\%$ error is shown.
The vertical line denotes  $\sigma=0.5$.}
\end{center}
\end{figure}

Since no mass function that we have considered is accurate for the
entire range of our data, we consider the possibility of an empirical
adjustment to the S-T function.  We insert a crude multiplicative
factor to the S-T function as follows, with $\delta_c$ $=$ 1.686 and
FOF $ll=$0.2 (Fig. 6):
\begin{equation}\label{sheth_tormen mod}
f(\sigma) = f(\sigma; {\rm S{\rm-}T})\bigg[exp[-0.7/(\sigma
[\cosh(2\sigma)]^5)]\bigg],
        \end{equation}
valid over the range of -1.7 $\leq
\ln\sigma^{-1} \leq$ 0.9.  The resulting function
is virtually identical to the S-T function
for all $-\infty \leq  \ln\sigma\leq0.4$.  At higher values
of $\ln\sigma^{-1}$, this function declines relative to the S-T
function, reflecting an underabundance of haloes that becomes greater
with increasing $\ln\sigma^{-1}$.  For -1.7 $\leq \ln\sigma^{-1} \leq$
0.5, Eqn. 9 matches our data to better than 10$\%$ for well-sampled
bins, while  for 0.5 $\leq \ln\sigma^{-1} \leq$ 0.9, where poisson
errors are larger, our data is matched  to roughly 20$\%$.  
We must
caution that though Eqn. 9 is a good fit to our data, it differs  from
the S-T function in the regime where poisson and cosmic errors are
highest, and where our  results are most prone to potential numerical
errors because of the steepness of the mass function.  Our results are
more robust in the low $\ln\sigma^{-1}$ regime.  Note that in the Eqn. 9 
fit, not all mass belongs to a halo, so Eqn. 3 is not valid.

\subsection{Friends-of-Friends (FOF) versus Spherical Overdensity (SO)}

Other authors have noted the advantages and disadvantages of the FOF
and SO algorithms (see Jenkins \etal 2001 and references therein).
The FOF method has the advantage that it can identify haloes of any
shape as long as their minimum local number density is at least
roughly $1/b^{3}$, and FOF is generally computationally cheaper than
SO.  However, FOF can sometimes spuriously link together haloes that
lie close together within a filament (see \eg  Governato \etal 1997).
In Fig. 9, we compare FOF mass functions of our simulation with the
corresponding SO mass functions for a range of redshifts.  Note that
the vertical axes are somewhat arbitrary for the z$=$10 and z$=$15
outputs as these were made from the z$_0=69$ version of the simulation
which had a somewhat suppressed mass function, which we discuss in
section 4.2.  The halo finders have excellent agreement at low
redshifts, with differences of $\simlt 10 \%$ over the range where the
mass function is well sampled.  Differences in the high mass bins are
due to a combination of different mass calculations for individual
selected clusters as well as offset mass bins.  The steep dropoff in
the SO mass function for low masses is due to a our exclusion of SKID
haloes of less than 64 particles as potential SO centers.  At high
redshifts, FOF $ll=$0.2 produces a substantially higher mass function
than FOF $ll=$0.164 or SO, which are similar to each other, implying
sensitivity to $ll$ for large $\ln\sigma^{-1}$, probably because the
mass function  is steep there, and thus sensitive to halo selection
criteria.  To verify that the discrepancy of FOF $ll=$0.2 with SO is
not due simply to our choice of using skid haloes as our initial SO
centers, we have included a high redshift (z$=$10) SO mass function
which uses FOF haloes as initial SO centers; our choice of centers
from which to grow our SO spheres has no effect on the SO mass
function.  A visual inspection in which halo members are ``marked'',
reveals that at high redshift, FOF $ll=$0.2 links together some
neighboring haloes connected by filaments, but FOF also identifies
some individual haloes (often of highly elongated shape) that are
missed by SO, so neither algorithm is ideal.  The overprediction of
the S-T function for rare objects worsens somewhat if we use SO
derived mass functions.  Using a linking length that varies with
redshift in an attempt to match the varying overdensity of virialized
haloes, would have little effect on our mass function, since at low
redshift the mass function is insensitive to $ll$, and at high
redshift $\Omega_{m}\simeq$1, implying $ll=$0.2 (Davis \etal 1985;
Lacel \& Cole 1994).  However, had we sampled large $\ln\sigma^{-1}$
at low redshift, adjusting $ll$ to match virial overdensity would
likely have a significant effect.

\begin{figure}
\begin{center}
\label{fig:sofof}
\epsfig{file=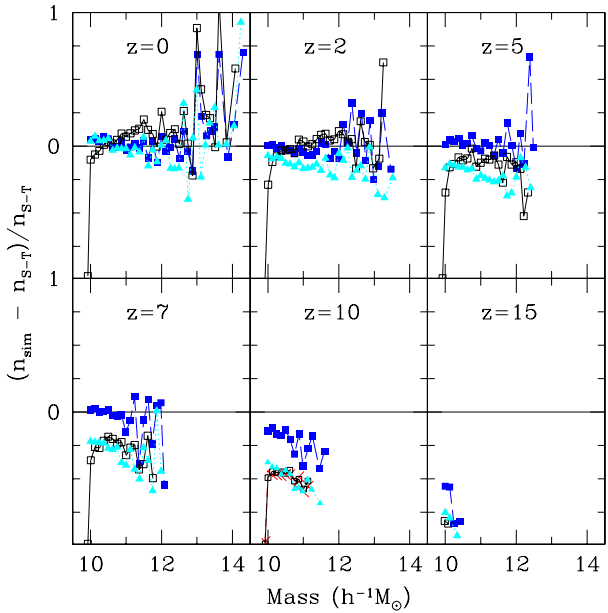, angle=0, width=\hsize}
\caption{Comparison of SO and FOF mass functions from our simulation.
Filled squares (connected by dashed lines) are for FOF $ll=$0.2
haloes; filled triangles (connected by dotted lines) are FOF
$ll=$0.164 haloes.  Open squares (solid lines) are SO haloes made with
SKID centers.  X's at z$=$10 are SO haloes made using FOF centers.  SO
overdensity criterion is  from Kitayama \& Suto 1996.  Note that the
$z=10$ and $z=15$ data lie too far below the S-T function in this plot
because it is from our z$_0=$69 run rather than  our z$_0=$139 run,
which we use for the rest of the high redshift results of this paper.}
\end{center}
\end{figure}

\subsection{Numerical Tests}

We check that the disagreement of S-T which appears at high redshift
is not a result of delayed halo collapse due to numerical errors that
can be caused by too large of a gravitational softening length.  We
make additional checks addressing potential numerical errors caused by
mapping particles with Zel`dovich displacements (Zel'dovich 1970) onto
a particle grid; an insufficiently high starting redshift could delay
collapse of the first haloes (\eg Jenkins \etal 2001).  If initial
conditions are set with some regions having overdensities high enough
to already be in the non-linear regime, then the linear Zel`dovich
mapping can not account for shell-crossing wherein mass piles up as it
flows toward overdensities.  The effects of either of these error
sources, if present, should have evolved away by lower redshifts,
since the tiny fraction of matter that is in haloes at such high
redshifts is soon incorporated into clusters or large groups.

\begin{table*}
\centering
\begin{minipage}{140mm}
\caption{N-body simulation parameters, including test runs}
\begin{tabular}{@{}llllllll@{}}
z$_0$  &  $r_{\rm soft}(h^{-1}{\rm kpc}$) & $\Theta$(z$>$7) &
n$_{replica}$ & z$_{evolved}$: & $\Theta$(2$<$z$<$7) & $\Theta$(z$<$2)
& \\ \hline 69 &   5.0  & 0.7 & 1 & 0 & 0.7 & 0.8 & Applied to z$<$7
results; SO vs. FOF test \\ 139 &  5.0  & 0.5 & 1 & 7 & -- & --  &
Applied to z$\geq$7 results \\ 139 &  5.0  & 0.5 & 2 & 7 & -- & --  &
Test \\ 69 &   2.5  & 0.7 & 1 & 7 & -- & --  & Test \\ 39 &   5.0  &
0.7 & 1 & 10 & -- & --  & Test \\ 279 &  5.0  & 0.5 & 2 & 7 & -- & --
& Test \\ &  & &  & & \\
\end{tabular}
\end{minipage}
\end{table*}

Table 1 lists our test runs, each of which consists of an  identical
432$^3$ particle volume with identical random waves, and is evolved to
z$=$7.  Fig. 10 shows the mass function for z$\simeq$7-15 for our low
softening test run, started from z$_0=$69, and plotted relative to the
S-T function along with our z$_0=$69, 5 $h^{-1}$kpc data.  Halving the
softening to 2.5 $h^{-1}$kpc has no effect on the mass function.
Fig. 11 shows the mass function for z$\simeq$7-15 for our initial
redshift test runs, all with 5 $h^{-1}$kpc softening.  Lowering the
initial redshift to 39 substantially reduces the number of high
redshift haloes, so we did not evolve the z$_0=$39 run to z$<$10.  The
z$_0=$139 run matches the z$_0=$279 run, indicating convergence, but
the z$_0=$69 run has a reduced mass function relative to the z$_0=$139
run at redshifts z$\simgt$12.  By z$=$7, however, the z$_0=$139 and
z$_0=$69 mass functions have converged, showing that evolving the
simulation over an expansion factor of $\simeq$10 from initial
conditions is sufficient for mass function measurements.  We
consequently derive our low redshift (z$<$7) results throughout this
paper from the z$_0=$69, 5 $h^{-1}$kpc simulation, and utilize the
z$_0=$139 run for our z$\geq$7 results.  We make an additional test of
z$\geq$7 cell opening angle,  which is used to determine how
accurately long range gravitational forces are to be approximated.  If
the cell opening angle is too large, then artificial net forces will
be incurred upon particles, which could cause ``spurious'' haloes to
form.  This effect is most likely to occur at high redshifts when
gravitational perturbations are small and force errors are
fractionally larger.  Decreasing the opening angle from $\Theta$=0.7
to $\Theta$=0.5, for our z$_0=$139 case, had no appreciable effect on
the mass function for z$\geq$7.  We test that  the number of replicas
used for our periodic boundaries, n$_r=$1, is  adequate.  With
z$_0=$139, increasing n$_r$ from 1 to 2  has no effect on the mass
function.  Additionally, to test that our box size is adequate, we
have verified that our z$=$0 mass function agrees with the mass
function from larger, lower resolution volumes where they overlap (not
included in Table 1).

\begin{figure}
\begin{center}
\label{fig:soft}
\epsfig{file=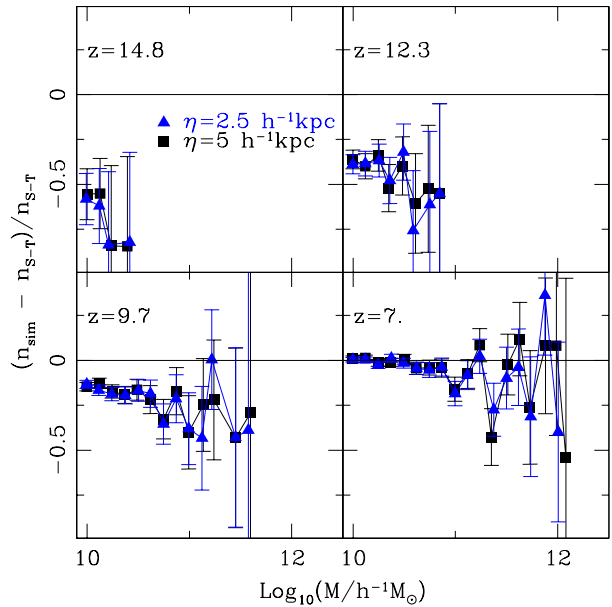, angle=0, width=\hsize}
\caption{The high redshift mass functions for runs with softening
lengths of 5 $h^{-1}$kpc and 2.5 $h^{-1}$kpc.  Both runs were started
from a redshift of 69.  }
\end{center}
\end{figure}

\begin{figure}
\begin{center}
\label{fig:z0}
\epsfig{file=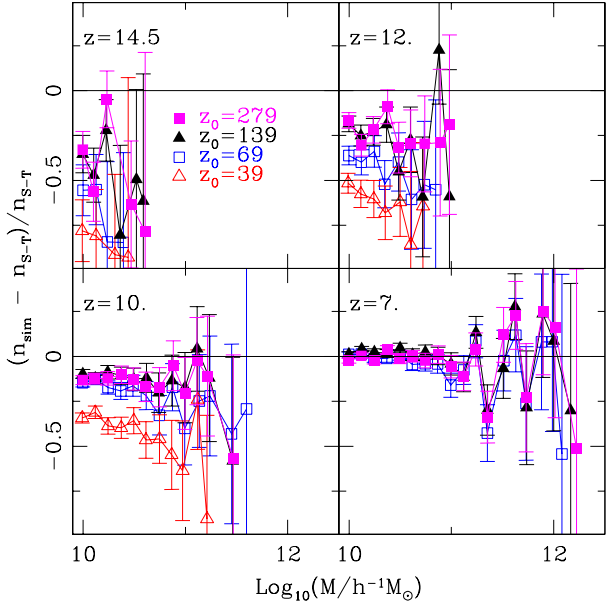, angle=0, width=\hsize}
\caption{The high redshift mass functions for initial redshifts of 39,
69, 139, and 279.  Each run has a softening of 5 $h^{-1}$kpc.  The
$z_0=$39 run was stopped at $z=$10.  }
\end{center}
\end{figure}

\section{Conclusions}

Our results extend to lower masses and higher redshifts than the
original empirical fit of the S-T mass function.  The range of masses
and redshifts over which the S-T mass function remains valid is quite
remarkable, and though it does begin to break down at redshift
$\simgt$ 10 in our results, no other function matches its range and
accuracy.  It is not well understood why the mass function can be
described so well by solely  $\sigma(M)$.   Lacey \& Cole 1994, in
simulations with scale free power spectra, found evidence that the
mass function depends on n$_{eff}$ as well as $\sigma$, though they
tested a  wider range in n$_{eff}$ than in more recent CDM
simulations.  The generally good agreement of our data with previous
work at much shallower  n$_{eff}$ implies that any such dependence is
weak, though it may be manifested  in our results where we differ from
the S-T function.  As n$_{eff}$ approaches -3, the growth of M$_*$
with time diverges, so any dependence of f($\sigma$) on n$_{eff}$ is
most likely to occur near that regime.  Simulations with finer mass
resolution will be able to test yet steeper values of n$_{eff}$.
Based on the apparent trend of the S-T function to overpredict halo
numbers for objects of greater rarity, we expect that the
overprediction of the S-T function may continue to worsen when even
higher values of $\ln\sigma^{-1}$, or equivalently, when more extreme
values of $M/M_*$ and redshift, are analyzed.  Theoretical work that
focuses on halo collapse in this high $\ln\sigma^{-1}$ range is needed
to produce more robust predictions.

Because the form of the mass function for low mass, low redshift
haloes closely resembles a power law, we are cautiously optimistic
that the S-T mass function will continue to provide a good match to
simulation data as lower values of $\ln\sigma^{-1}$ are modeled,
though  this extrapolation will likely breakdown where n$_{eff}$
approaches -3.  Simulations that model higher particle numbers
(leading to higher redshifts), are needed to extend the known range of
the mass function of dark matter haloes.  The accuracy of the S-T
function for low mass haloes out to high redshifts has important
implications for a number of astrophysical problems.  Evolution of the
mass and luminosity functions down to dwarf scales permits comparison
with surveys, providing important cosmological tests, and allows
calculations of merger histories and star formation histories of
galaxies, groups, and clusters.  Our results verify that the S-T
function is accurate over the entire evolutionary range (for which
progenitors or descendants are observable) of lyman break galaxies and
groups of galaxies.  Assuming a weak dependence on n$_{eff}$,  the
redshift invariance of the mass function implies that extrapolation of
mass functions should be reliable for combinations of masses and
redshifts that cannot presently be simulated, as long as only values
of $\ln\sigma^{-1}$  that have been verified by simulation are
considered.   The number densities of low mass ($< 10^{10} \msun$)
haloes at high redshift (z $\sim$ 10), needed for studies of
reionization, or galaxy formation (\eg Haiman 2002), should be well
described by the S-T function since although they lie below our mass
range, they are within our range of simulated $\ln\sigma^{-1}$.  For
such  extrapolation to be accurate down to indefinitely small masses,
all mass would have to be in dark matter haloes of some mass or else
low mass haloes would be overpredicted.

Extrapolation of mass functions to large values of $\ln\sigma^{-1}$
that have yet to be verified by simulations, however, are likely to be
significantly in error, as suggested by the trend of increasing
overprediction by the S-T function for high values of
$\ln\sigma^{-1}$.  Though our results only reach $\simeq$4$\sigma$
density peaks, there is a trend for the S-T function to increasingly
overpredict the mass function for increasing $\sigma$ beginning at
$\simeq$3$\sigma$.   The discrepancy with the S-T function for rare
objects has significant implications for studies that make use of such
rare objects as a cosmological probe.  For example, the number density
of high redshift (z $\simeq$ 6) QSOs, which are thought to be hosted
by haloes at 5$\sigma$ peaks in the fluctuation field (Haiman \& Loeb
2001; Fan \etal 2001), are likely to be overpredicted by at least a
factor of 50$\%$.  Some uncertainty is also introduced for studies
employing the abundance of the highest redshift clusters to probe
cosmological parameters (\eg Robinson, Gawiser, \& Silk 2000 and
references therein).

\subsection{Summary}
In summary, we have utilized high resolution simulations to derive the
mass function of dark matter haloes over an extended range in both
mass and redshift over previous work.  We find that the S-T mass
function holds up exceptionally over more than 10 orders of magnitude
of $M/M_*$.  For -1.7 $\leq \ln\sigma^{-1} \leq$ 0.5, the S-T mass
function is an excellent fit to our data, but begins to overpredict
haloes for ln$(\sigma^{-1}) \simgt$ 0.5, or $M/M_*$ $\gtsima$ 10$^{6}$
in our volume, reaching a $\sim$50$\%$ discrepancy by ln$(\sigma^{-1})
\simeq$ 0.9, corresponding to $M/M_*$ $\sim$ 10$^{9}$ at z$\simeq$15
in our volume.  We offer an empirical adjustment for the high
ln$(\sigma^{-1})$ portion of the S-T mass function.  Our results
confirm the redshift invariance of the mass function.

\section*{Acknowledgments}
We graciously thank the referee, Adrian Jenkins, for extremely
helpful suggestions for improvements to this paper.
This work was supported, in part by NASA training grant NGT5-126.  Our
simulations were performed on the Origin 2000 at NCSA and NASA Ames,
the IBM SP4 at the Arctic Region Supercomputing Center (ARSC), and the
NASA Goddard HP/Compaq SC 45.  We thank Chance Reschke for dedicated
support of our computing resources, much of which were graciously
donated by Intel.

{}

\label{lastpage}

\end{document}